\documentclass{aa}

\usepackage{graphicx}
\usepackage{natbib}
\usepackage{psfrag}
\usepackage{verbatim}
\graphicspath{{./figures/}}

\newcommand{\g}[1]{{#1}}

\begin{document}

\title{Five new very low mass binaries
\thanks{Based on observations obtained at the Canada-France-Hawaii
Telescope (CFHT) which is operated by the National Research Council
of Canada, the Institut National des Sciences de l'Univers of the
Centre National de la Recherche Scientifique of France, and the
University of Hawaii; and observations made at the European Southern
Observatory, Paranal, Chile through proposal 075.C-0733(A) and GTO NACO.}}

\author{G. Montagnier \inst{1,2}
\and D. S\'egransan \inst{2}
\and J.-L. Beuzit \inst{1}
\and T. Forveille \inst{3,1}
\and P. Delorme \inst{1}
\and X. Delfosse \inst{1}
\and C. Perrier\inst{1}
\and S. Udry \inst{2}
\and M. Mayor\inst{2}
\and G. Chauvin\inst{5}
\and A.-M. Lagrange\inst{1}
\and D. Mouillet \inst{4}
\and T. Fusco \inst{6}
\and P. Gigan \inst{7}
\and E. Stadler \inst{1}
}

\institute{ Laboratoire d'Astrophysique de Grenoble,
                 BP53,
                 F-38041 Grenoble Cedex,
                 France
\and
                Observatoire de Gen\`eve,
                51, chemin des Maillettes
                CH-1290 Sauverny,
                Switzerland
\and
                Canada-France-Hawaii Telescope Corporation,
                65-1238 Mamalahoa Highway,
                Kamuela, HI 96743, Hawaii
                U.S.A.
\and
                Laboratoire d'Astrophysique de Toulouse et Tarbes, BP826 F-65008 Tarbes Cedex, France
\and
                European Southern Observatory, Casilla 19001, Santiago 19,
                Chile
\and
                ONERA-DOTA, BP72, 29 avenue de la Division Leclerc, F-92322
                Ch\^atillon Cedex, France
\and
                Laboratoire d'\'Etudes Spatiales et d'Instrumentation
                Astrophysique, F-92195 Meudon Cedex,
                France
}

\date{Received}

\abstract{We report the discovery of companions to 5 nearby late M dwarfs
($>$M5), LHS1901, LHS4009, LHS6167, LP869-26 and WT460, and we \g{confirm}
that the recently discovered mid-T brown dwarf companion to SCR1845-6357 is
physically bound to that star.
These discoveries result from our adaptive optics survey of all M dwarfs
within 12~pc. The new companions have spectral types M5 to L1, and
orbital separations between 1 and 10~AU. They add
significantly to the number of late M dwarfs binaries in the
immediate solar neighbourhood, and will improve the multiplicity
statistics of late M dwarfs. The expected periods range from 3 to 130
years. Several pairs thus have good potential for accurate mass 
\g{determination} in this poorly sampled mass range.

}

\maketitle

\section{Introduction}
\g{ Stellar masses can only be measured empirically by
determining the orbits of multiple stars, and stellar multiplicity 
is a key parameter for several important astrophysical
issues:  models of stellar formation and
of the subsequent dynamical evolution of multiple systems should reproduce
the multiplicity fraction of every stellar class, as well as the
distributions of their orbital elements, and a good handle on stellar
multiplicity is crucial to correcting observed luminosity functions for
unresolved components. }

The stellar multiplicity statistics is now well established for solar
neighbourhood G \& K dwarfs \citep{duquennoy1991, halbwachs2003}
and is converging for M0-M5 dwarfs (\citeauthor{delfosse1999},
\citeyear{delfosse1999}, and in prep.). The binary fractions clearly
decrease with mass, from 57~\% for G~dwarfs \citep{duquennoy1991},
to 29$\pm$5\% for M0-M5 dwarfs (Delfosse et al., in prep). That fraction 
apparently drops further for late M dwarfs to 9$^{+4}_{-3}\%$ beyond 
3~AU \citep{siegler2005}, compared to $18\pm 4\%$ for early-M dwarfs 
over the same separation range.
The total binary fraction of late-M dwarfs is however uncertain, because
current observations have limited sensitivity to significantly closer
binaries, which have been suggested to be fairly numerous
\citep{jeffries2005}.
We are therefore observing with adaptive optics (AO) a well defined, nominally
volume limited, sample of  over 40 
late-M dwarfs ($>$M5) within 12~pc (Delfosse et al. in prep.). 
The close limiting distance of 
that sample ensures optimal linear resolution, and
provides good overlap with the sensitivity range of a \g{planned}
search for spectroscopic binaries.
In this paper we present 5 new companions to late-M primaries from that
sample, and confirm that the mid-T brown dwarf companion to SCR1845-6357
\citep{biller2006} is physical. Section 2 describes our observations and
data analysis, while Section 3 briefly discusses the physical parameters
of the new binary  systems.

 \section{Observations and data reduction}

\begin{figure}
\begin{tabular}{cc}
\includegraphics[width=4.1cm]{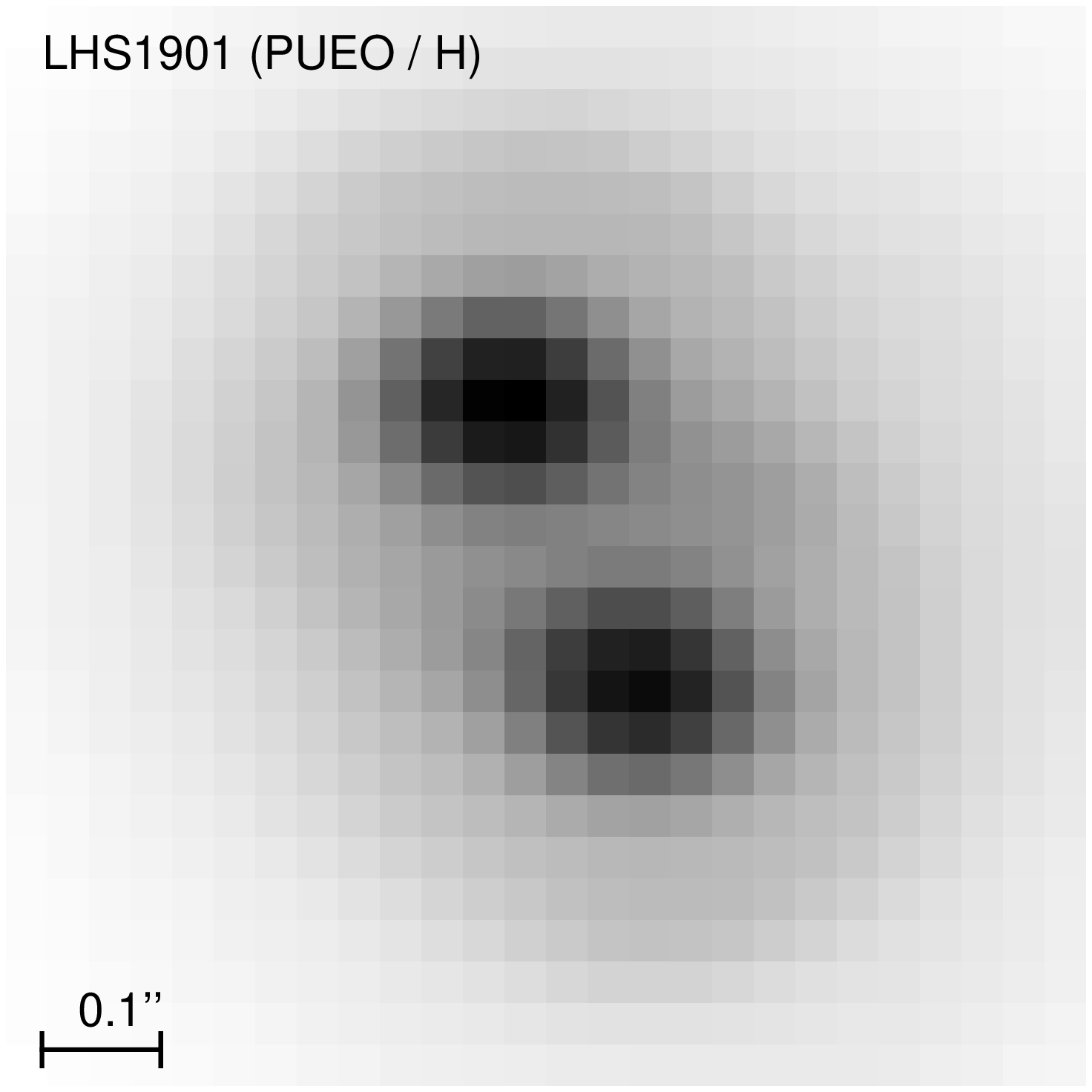} &
\includegraphics[width=4.1cm]{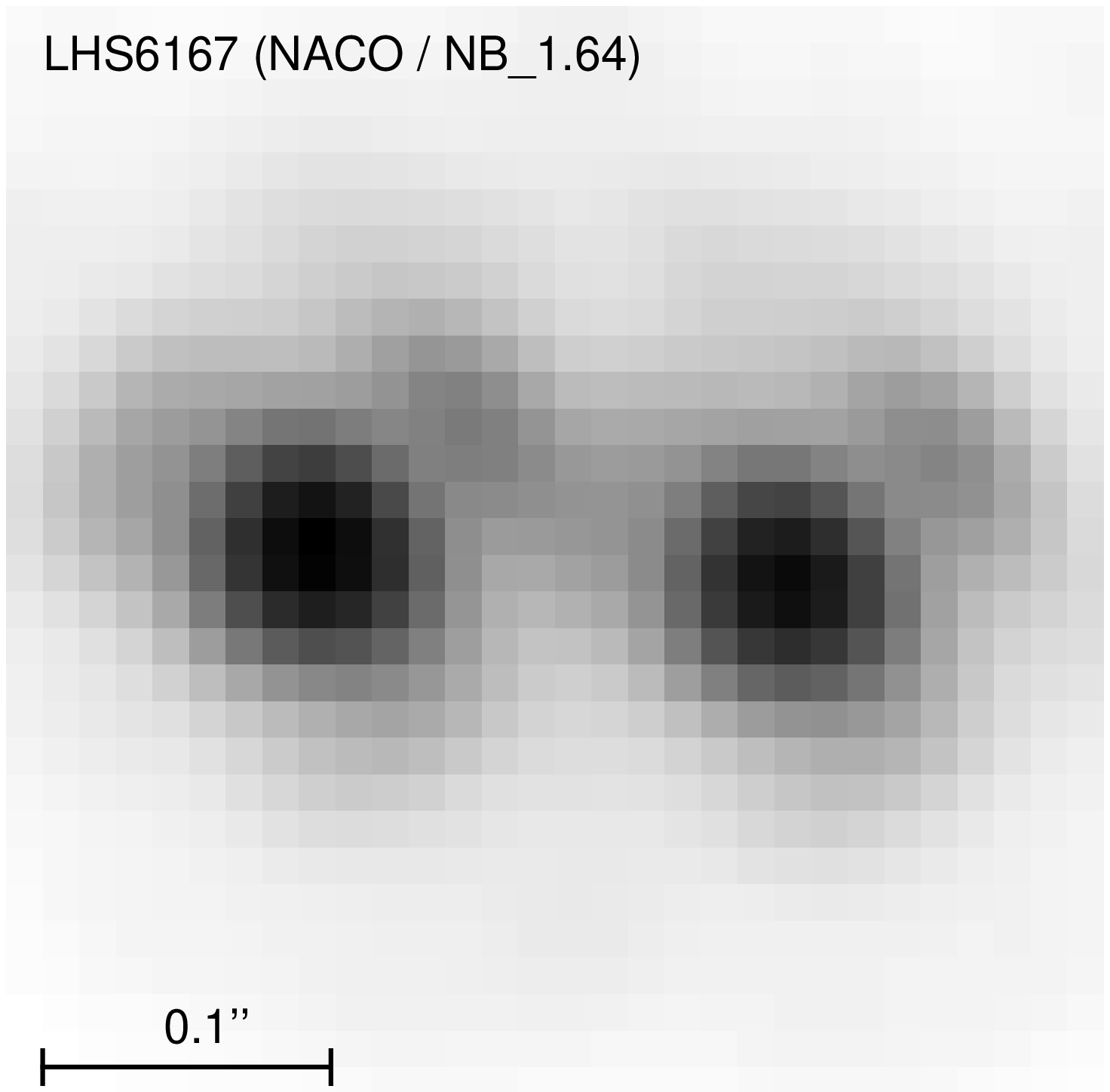} \\
\includegraphics[width=4.1cm]{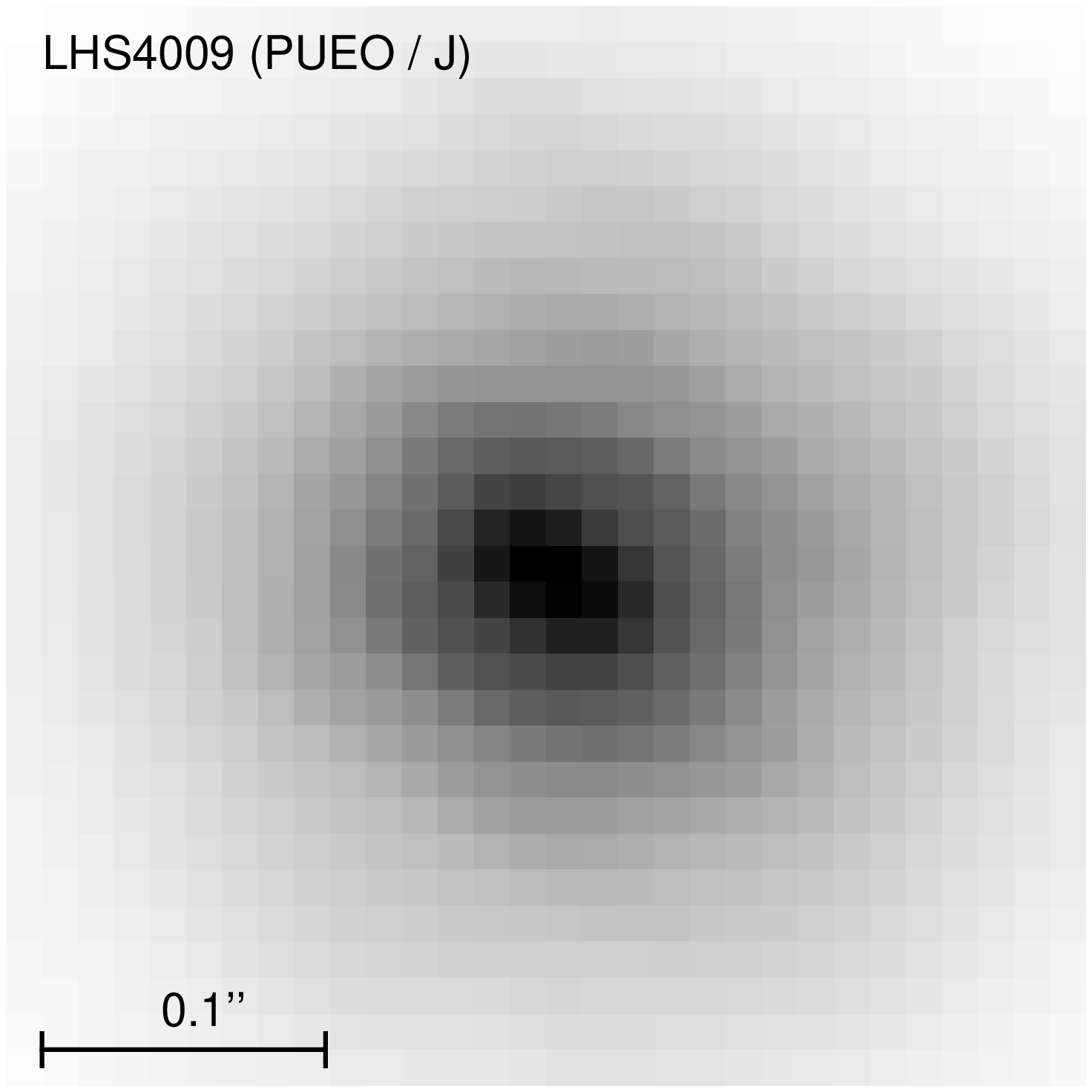} &
\includegraphics[width=4.1cm]{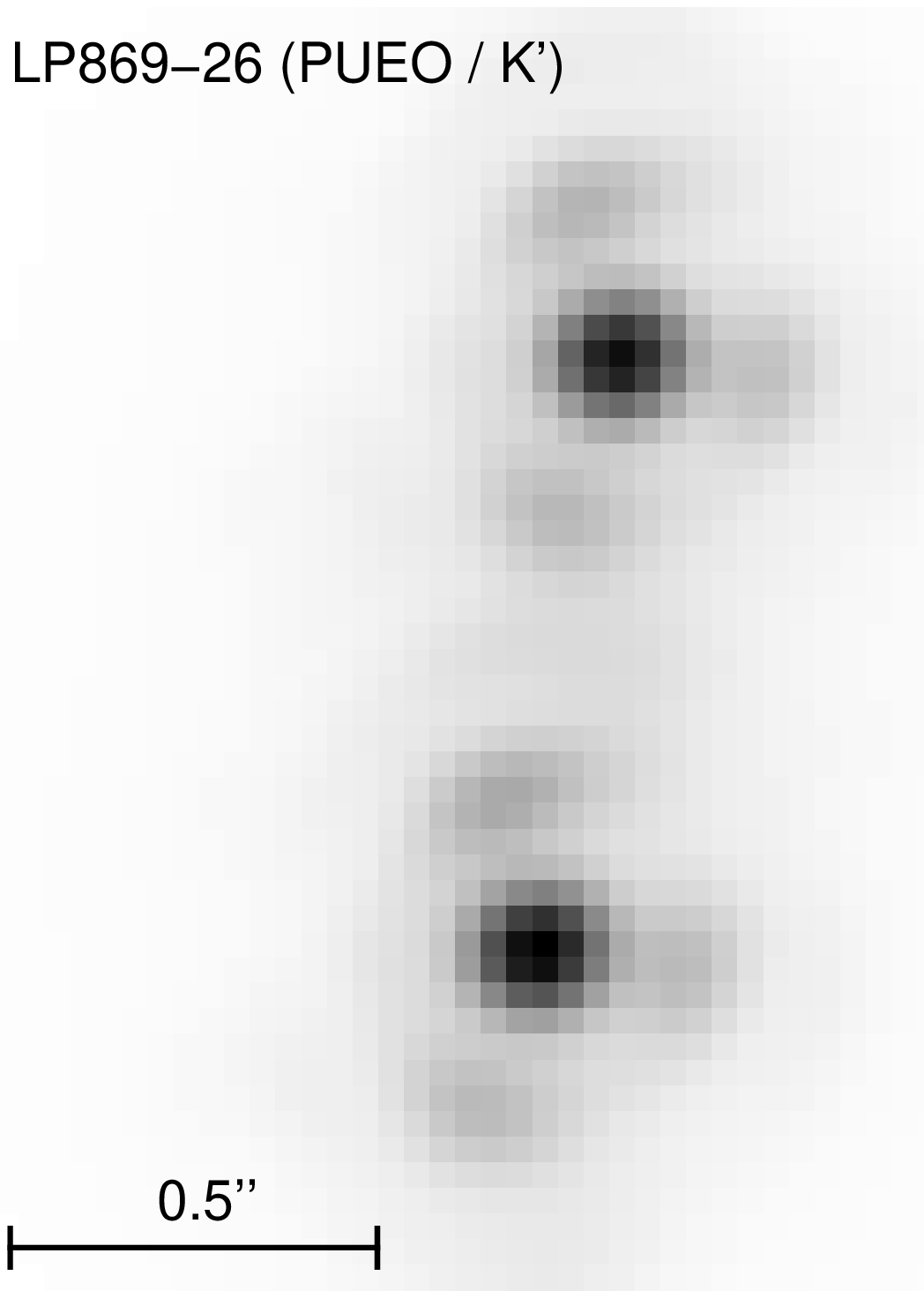} \\
\includegraphics[width=4.1cm]{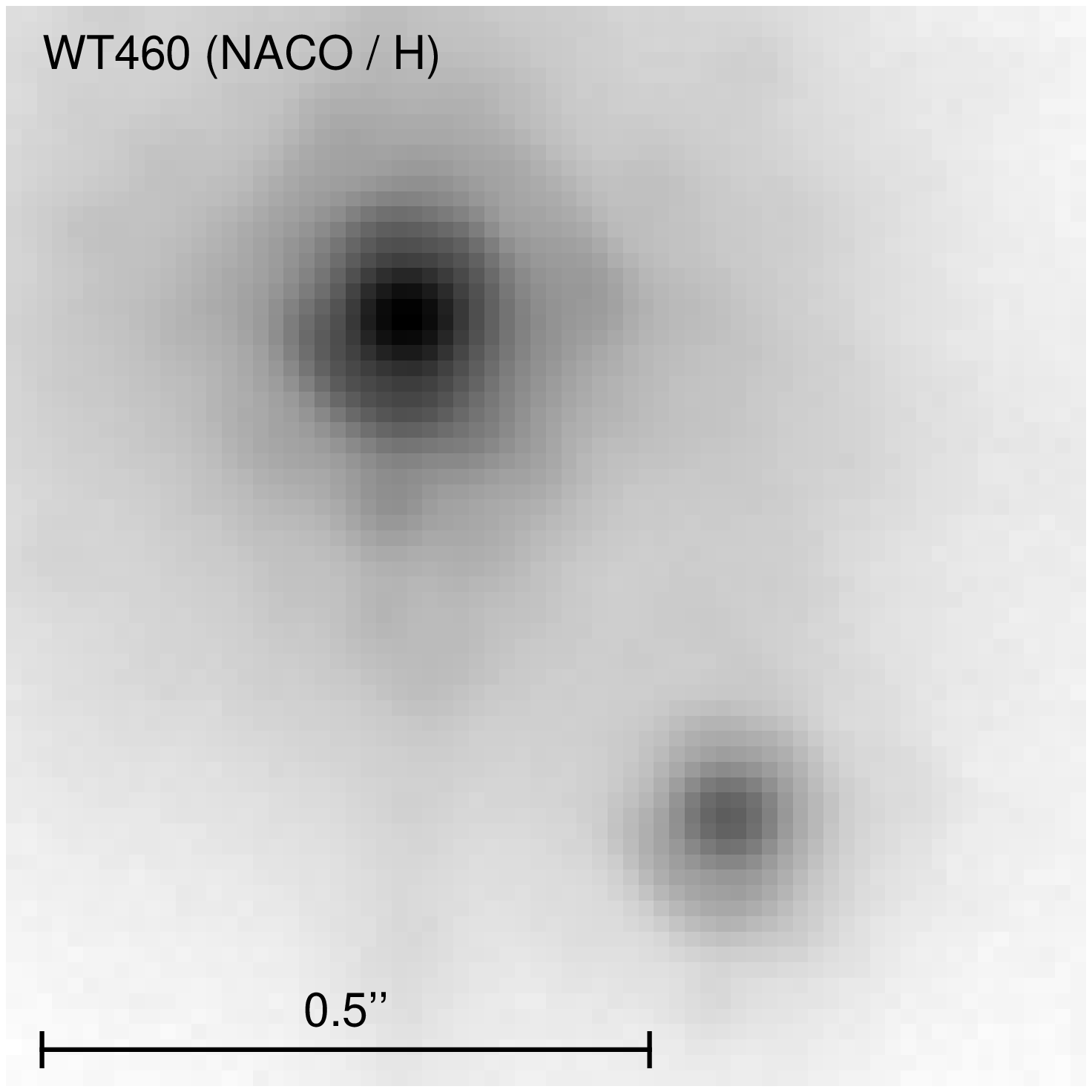} &
\includegraphics[width=4.1cm]{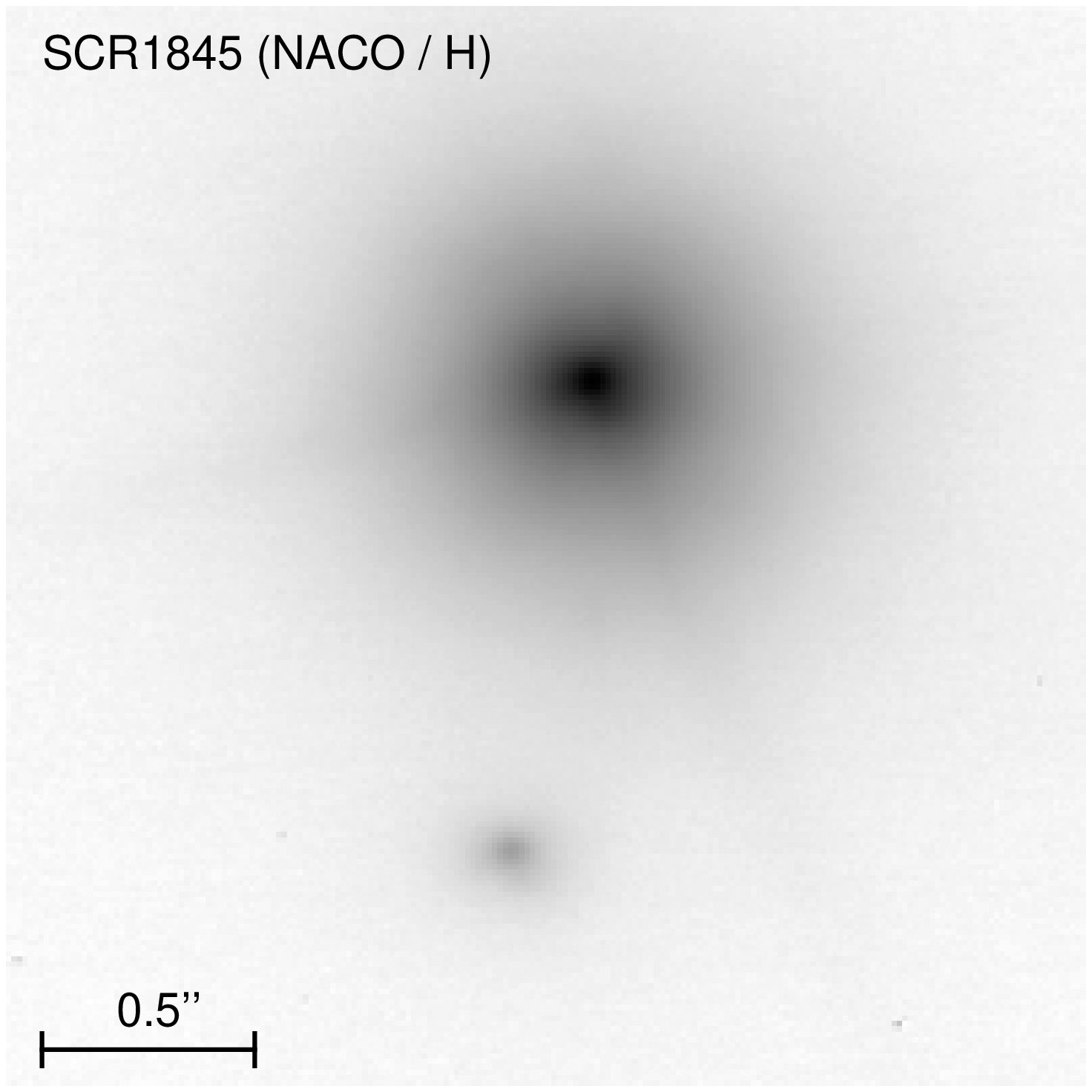} \\
\end{tabular}
\caption{AO images of the six resolved binaries.
The images of LHS 1901, LHS 4009, LP 869-26, LHS 6167 are displayed
with a linear stretch. Those of SCR~1845-6357 and WT 460 use
a logarithmic stretch. North is up and East is to the left.
}

\label{fig:images}
\end{figure}

LHS 1901, LHS 4009, and LP 869-26 were observed at the
Canada-France-Hawaii Telescope (CFHT), using the Adaptive
Optics Bonnette \citep{rigaut1998} and the KIR camera, with observing
procedures documented in \cite{beuzit2004}. LHS~6167, SCR~1845-6357 
and WT~460 were observed with the NACO instrument 
\citep{rousset2003,lagrange2003,lenzen2003} on VLT UT4 (ESO Very Large 
Telescope, Paranal Chile). We used the IR wavefront sensor of NAOS,
which produces the best corrected images for such faint red stars.
The NACO observation sequences consisted of several exposures
at  7 random positions within a 5~$\arcsec$ jitter box. The total
integration time was kept $>$5~minutes, to average the
residual speckle noise and increase  the signal to noise ratio of the
images. The faintest companion, SCR~1845-6357B ($\Delta$(K)=5.1), is 
however easily visible in the individual raw images, illustrating the 
impressive performances of NACO in its classical imaging mode.

The images were reduced with the ECLIPSE package \citep{devillard1997},
following standard steps for near-IR imaging: masking of bright and dead
pixels, flat-fielding, subraction of the sky background as estimated
from a median of the jittered images (iteratively masked for astronomical
sources), and stacking after the images had been aligned through
cross-correlation. The flat-field map was derived from images of the
illuminated dome at CFHT, and from images of the sunset sky for
the VLT observations. We reconstructed the long-exposure PSF associated 
with each CFHT image from the data of the wavefront sensor and the 
deformable miror commands \citep{veran1997}. For the NACO observation of 
SCR~1845-635 we use the primary star as the reference PSF, since the two
components are well
separated. For the other NACO targets we use a single star observed
during the same night. We extracted the coordinates and intensities of
the two stars by least-squares fitting two scaled PSFs to each 
image \g{\citep{veran1999}}.
Astrometric calibrations derived from observations of the standard Orion 
field \citep{mccaughrean1994} and of wide Hipparcos binaries then provided
the position angle and separation.
Figure~\ref{fig:images} displays one reduced image for each target, and
Table~\ref{tab:reduction_results} summarizes the extracted parameters.

\begin{figure}
\center
\begin{tabular}{c c}
\includegraphics[width=4.1cm]{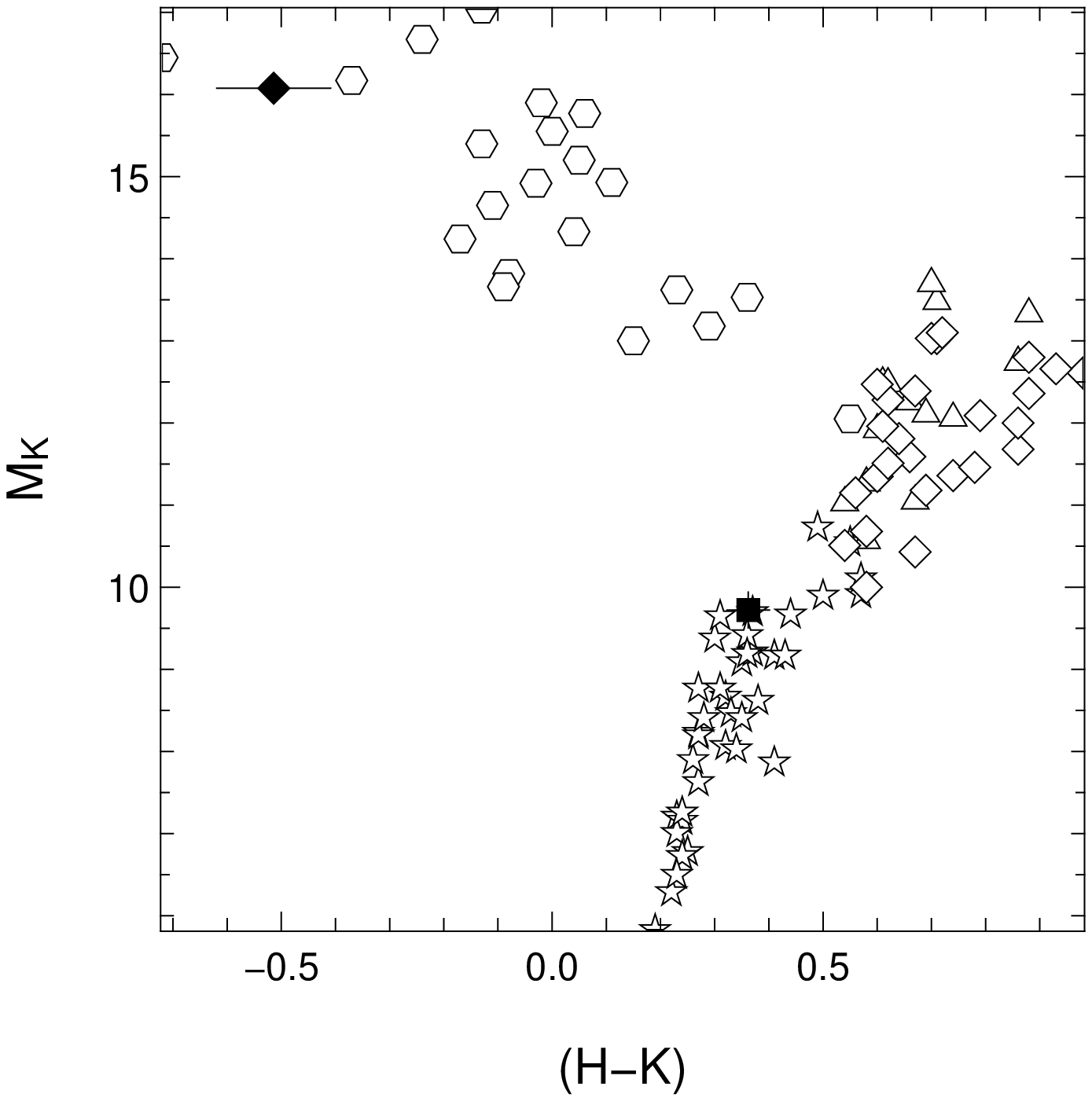} &
\includegraphics[width=4.1cm]{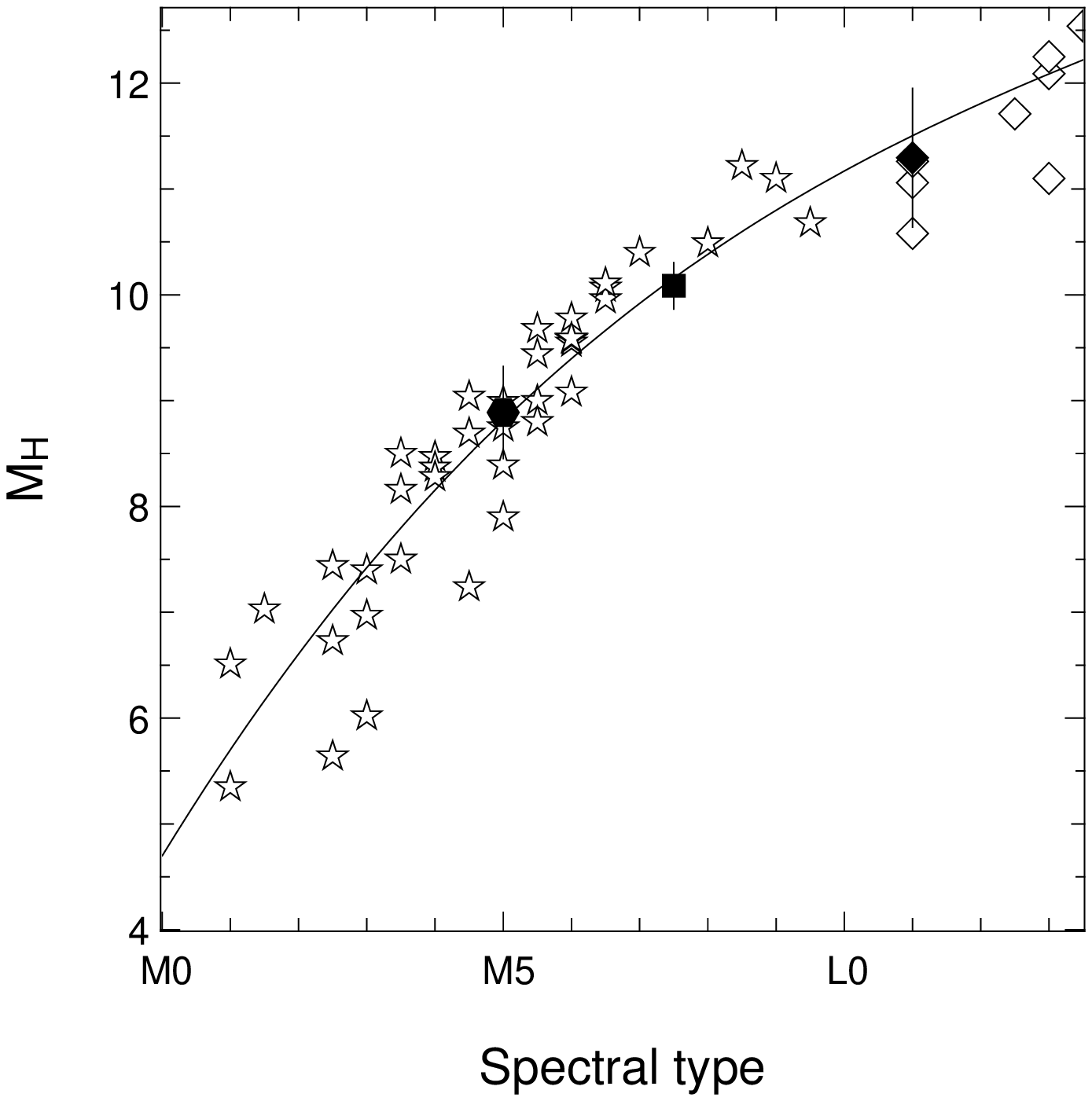}\\
\end{tabular}
\caption{Left:Absolute $K$ magnitude vs $H-K$. \g{The full hexagon is LHS1901B,
the full square is LHS6167B and the full diamond is WT460B.}
Right: Absolute $H$ magnitude vs spectral type. The plain line is a
polynomial
fit of the reference observational data points for the function
$\textrm{M}_\textrm{H}(\textrm{S.Type})$. \g{The full diamond is SCR1845-6357B and
the full square is LHS1901B.}
For both figures: 
empty stars are M-dwarfs from \citet{leggett2000} and \citet{leggett2002},
empty triangles are L-dwarfs from \citet{leggett2002}, empty diamonds
are L-dwarfs from \citet{knapp2004}, and hexagons are T-dwarfs from
\citet{knapp2004}.
}
\label{fig:diagramms}
\end{figure}

We computed magnitudes for the individual components from the 2MASS
photometry of the systems \citep{skrutskie2006} and our adaptive
optics flux ratio. \g{For SCR1845-6357 we used synthetic photometry
from apropriate spectra from S.K. Leggett's on-line spectral database
to correct the NB\_1.64 flux ratio to the broader K band. 
(K$-$NB\_2.17$=0.23 \pm 0.07$ for a M8.5-Mid-T system).}
We then approximately corrected the photometric
distances of the systems (except SCR1845-6357, which has an accurate
trigonometric parallax) for the light of the previously unrecognized
components, and we estimated \g{indicative} spectral types from plots of
absolute magnitude versus spectral type and colour versus spectral
type, derived from \citet{leggett2000,leggett2002} and
\citet{knapp2004}. \g{Figure \ref{fig:diagramms} shows one of these
diagrams.} We derived masses
using the observational mass-luminosity (M-L) relations of \cite{delfosse2000}
up to spectral type M7, \g{the DUSTY model of \citet{chabrier2000}
at 1 to 5 billion years for spectral types M8 to L}, and the COND
models of \citet{baraffe2003} for the mid-T SCR1845B. The estimated 
semi-major axes are 1.35 times the observed separation, a 
statistical factor between the semi-major axis of a visual binary
and its projected separation at discovery from \citet{duquennoy1991}. 
We computed approximate orbital periods from the mass and 
semi-major axis. They are uncertain by a factor of ${\sim}3$, 
dominated by the \g{(factor of ${\sim}2$)} semi-major axis uncertainty 
(Tab.~\ref{tab:systems}).

\begin{table*}[htbp]
  \caption{Systems with new low-mass companions.\g{$^{(a)}$ m$_K$ from the 
2MASS catalogue \citep{skrutskie2006}. $^{(b)}$ Henry et al., in prep.}
}
  \label{tab:systems}
  \centering
  \small
  \begin{tabular}{c c c c c c c c c}
    \hline\hline
    Name      & m$_K$  &  Distance         & Spectral  & Spectral  & Mass A & Mass B & Period & semi-major axis\\
              &$^{(a)}$& (pc)              & type A    & type B    & ($\textrm{M}_{\odot}$) & ($\textrm{M}_{\odot}$) & (yr) & (AU) \\
    \hline
    LHS1901  & 9.12   & $ 11.0 \pm 1.1 $   & M6.5     & M7.5     &  $0.085 \pm 0.02$ &  $0.082 \pm 0.02$ & 20 & 4\\
    LHS4009  & 8.31   & $ 14.6 \pm 4.4 $   & M5       & M5       & $0.165 \pm 0.05$  & $0.155 \pm 0.05$  & 3  & 1.3\\
    LHS6167  & 7.73   & $ 10.0 \pm 2.0 $   & M5       & M5       & $0.14 \pm 0.05$   & $0.13\pm 0.05$   & 6  & 2\\
    LP869-26 & 8.26   & $13.1 \pm 2.6$     & M5.5     & M6.0     & $0.15 \pm 0.05$   & $0.14 \pm 0.05$   & 130& 15\\
    WT460    &  8.62  & $11.6 \pm 3.5$     & M6       & L1       & $0.12 \pm 0.04$   & 0.070-0.080 & 40 & 6\\
    SCR1845  & 8.51   &$3.85 \pm 0.02^{(b)}$&  M8.5   &  mid-T      & 0.075-0.080 & 8-65$\textrm{M}_{Jup}$ & 35 & 6  \\
    \hline
\end{tabular}
  \end{table*}

\begin{table*}[htbp]
  \caption{AO measurements of the new low-mass companions.
}
  \label{tab:reduction_results}
  \centering
  \small
  \begin{tabular}{c c c c c c c c c c c c}
    \hline\hline
    Name & $\rho$    & $\theta$ & $\Delta m$ & Date & Filt.  & Intrument & Strehl & reference\\
         & (\arcsec) & (\degr)  &            &      &        &           & (\%)   & star     \\
    \hline
    LHS1901  & $0.275 \pm 0.005$ & $208.0 \pm 0.5$   & $0.13 \pm 0.03$   & 08 Jan 2004 & K'       & PUEO & 37 & LP205-49 \\
             & $0.204 \pm 0.005$ & $215.0 \pm 0.5$   & $0.07 \pm 0.03$   & 27 Apr 2005 & K'       & PUEO & 55 & LP423-31 \\
             & $0.174 \pm 0.005$ & $219.6 \pm 0.5$   & $0.14 \pm 0.05$   & 14 Oct 2005 & H        & PUEO & 36 & LP205-49\\
    LHS4009  & $0.068 \pm 0.007$ & $238 \pm 3$       & $0.15 \pm 0.1$    & 14 Oct 2005 & J        & PUEO & 12 & HIP482\\
             & \g{$0.066 \pm 0.007$} & \g{$250 \pm 10$}      & \g{$0.14 \pm 0.1$}    & \g{14 Oct 2005} & \g{K'}       & \g{PUEO} & \g{47} & \g{L707-74}\\
    LHS6167  & $0.076 \pm 0.001$ & $82.4 \pm 0.3$    & $0.12 \pm 0.01$   & 12 Sep 2003 & NB\_1.64 & NACO & 45 & Gl317\\
             & $0.172 \pm 0.001$ & $265.8 \pm 0.1$   & $0.13 \pm 0.01$   & 01 May 2005 & NB\_1.64 & NACO & 58 & Gl680\\
    LP869-26 & $0.813 \pm 0.005$ & $354.7 \pm 0.3$   & $0.08 \pm 0.01$   & 03 Jul 2004 & K'       & PUEO & 53 & LP869-19\\
             & $0.828 \pm 0.005$ & $353.1 \pm 0.1$   & $0.08 \pm 0.01$   & 14 Oct 2005 & K'       & PUEO & 25 & LP869-19\\
    WT460    & $0.511 \pm 0.001$ & $212.6 \pm 0.1$   & $2.47 \pm 0.05$   & 01 May 2005 & H        & NACO & 54 & primary\\
\hline
    SCR 1845       & $1.176 \pm 0.001$ & $170.22 \pm 0.08$ & $4.43 \pm 0.05$   & 01 May  2005 & H        & NACO  & 42 & primary\\
                   & $1.176 \pm 0.001$ & $170.22 \pm 0.08$ & $5.14 \pm 0.05$   & 01 May  2005 & NB\_2.17 & NACO  & 30 & primary\\
\citet{biller2006} & $1.170 \pm 0.003$ & $170.20 \pm 0.13$ & $4.19^{+0.31}_{-0.26}$ & 28 May 2005 & H    & NACO+SDI & &        \\
\hline

  \end{tabular}
\end{table*}

\section{Discussion}

In spite of figuring in the Luyten Half Second catalogue
\citep{luyten1979}, LHS~1901 was only recently identified as a nearby
star by \citet{reid2003}, who derived a spectral type of M6.5 and a
photometric distance of $8.0 \pm 0.8~\textrm{pc}$ (assuming
a single star). The new companion is clearly bound to LHS~1901: that
star has a $0.73\arcsec / \textrm{yr}$ proper motion \citep{salim2003},
and the separation changes by only ${\sim}0.1\arcsec$ over 1.5~yr.
Correcting the \citet{reid2003} photometric distance
for the light of the new companion moves the system out to 11.1~pc.
At that distance, the absolute magnitude of the new companion corresponds
to an M7.5 star, and the \citet{delfosse2000} K-band M-L relation 
predicts  masses of $0.085~\textrm{M}\odot$
and $0.082~\textrm{M}\odot$. The expected period
of approximately 20~years is  consistent with the ${\sim}0.1\arcsec$
observed motion over 1.5~year, and gives good prospects for accurate mass
measurements within a realistic time.\\
\\
\citet{hawley1996} first added LHS~4009 to the solar neighborhood
inventory, finding an M5  spectral type and a 10.6~pc spectrophotometric 
distance. Correcting for the light of the companion moves the system out
to 14.6~pc. The binary is marginally resolved in our J-band CFHT image, 
and just elongated in two K-band images at the J-band epoch and in August 
2001. \g{The August 2001 image is insufficiently resolved to derive 
quantitative parameters, since the binary was then even closer.}  
The $\sim$3~yr period
makes the system an excellent candidate for a very accurate mass 
determination.\\
\\
\citet{reid2002} derived a photometric distance for LHS~6167 and
identified it as a solar neighbour. \citet{scholz2005} derived a
spectral type of M5 and a more acurate spectrophotometric distance.
The relative motion of the components over 1.5~yr is only 
$0.25\arcsec$, consistent with the $\sim$6~yr orbital period, 
and excluding a background object given the $0.440\arcsec / \textrm{yr}$
proper motion \citep{salim2003}. The system has excellent potential
for accurate masses.\\
\\

LP 869-26 (also NLTT 48178) figures in the New Luyten 2-Tenths Catalogue,
and was recognized as a close neighbour by \citet{reid2003}.
\cite{scholz2005} determined an M5 spectral type, and a more accurate 
spectrophotometric distance. The $0.349\arcsec / \textrm{yr}$ proper motion 
of LP 869-26 \citep{salim2003} and the 1.25~yr interval between our two observations
ensure that the pair is gravitationally bound. Correcting for the light
of the 
new companion moves the system out to 13~pc.\\
\\
The WT~460 high proper motion star was discovered by \citet{wroblewski1991},
and identified as a member of the immediate solar neighbourhood by
\citet{patterson1998}, who used VRI photometry to derive an 11~pc photometric
distance. That photometry corresponds to an $\sim$M6 spectral type. Our adaptive
optics image shows a faint companion, with a magnitude which at
that distance corresponds to an early-L dwarf. Since we observed WT460
at a single epoch, we must rely on external input to demonstrate that the
pair is bound.
The high proper motion of WT~460 ($0.77\arcsec/\textrm{yr}$,
\citeauthor{wroblewski1991}, \citeyear{wroblewski1991}) ensures that
it does not interfere with examination of its May 2005 position
($14{^h}12{^m}59{\fs}56$ $-41{\degr}32{\arcmin}20{\farcs}9$) in the
DSS images. That position is a blank field in the Blue (May 1$^{st}$ 1979),
Red (June 23$^{rd}$ 1995), and I (June 17$^{th}$ 1994) images, with
approximate limits of
$B \sim 21.5$, $R \sim 20.8$ and $I \sim 19.5$. This ensures that the
companion either shares the proper motion of WT~460, or is extremely red,
with $\textrm{B-H} > 10$, $\textrm{R-H} > 9$ and $\textrm{I-H} > 8$.
A background star would thus have to be a T dwarf, which at
H=11.5 would then be the second brightest known to date. The probability
of finding such a rare object in the background of our small sample is
vanishingly small, demonstrating that the two objects form a
common proper motion pair. We also examined the April 1999
2MASS images, where WT~460 is $4.7\arcsec$ away from its May 2005 position.
A $\Delta({\mathrm H})$=2.5 background object at a $4.7\arcsec$
separation would be obvious, and the star is unresolved. WT460B is
thus a bound companion, with a photometry consistent with spectral
type $\sim$L1.\\ 
\\
SCR 1845-6357 (hereafter SCR1845) is a very recent addition to the solar
neighborhood inventory, discovered by \citet{hambly2004}. This M8.5 star
\citep{henry2004} now has an accurate trigonometric parallax
(Henry et al., in prep), 
 and is the 24$^{th}$ closest stellar
system to the Sun\footnote{see: http://www.chara.gsu.edu/RECONS/TOP100.htm}.
Even more recently, \citet{biller2006} identified a very faint
companion, for which they obtained a mid-T spectral type from
intermediate-band photometry. We present here an independent observation
of that companion, which demonstrates that systems with the observational
characteristics of SCR 1845 ($\Delta{\mathrm{H}}\sim 4.5$ at
$\sim 1\arcsec$) 
are easily accessible to standard imaging
with NACO \g{and its IR wavefront sensor}, 
and do not require the added complication of the differential
mode used by \citet{biller2006}. We use our detection to establish, if
need there be, that the pair is gravitationally bound, and to derive
improved broad-band photometry. Thanks to the very large proper motion
of SCR1845 ($2.64\arcsec/\textrm{yr}$, \citeauthor{deacon2005},
\citeyear{deacon2005}), the one month baseline between our May 1$^{st}$
2005 observation and the May 28$^{th}$ \citet{biller2006} measurement
is amply sufficient to establish common proper motion: while
SCR1845 moved by $0.195\arcsec$ between the two dates,
the separation of the pair changed by only $0.006{\pm}0.003\arcsec$.
From its position in the M$_{\mathrm K}$ vs H$-$K H-R
diagram, \g{SCR1845B is a mid-T dwarf}.

\bibliographystyle{aa}
\bibliography{mybiblio}

\begin{thebibliography}{33}
\expandafter\ifx\csname natexlab\endcsname\relax\def\natexlab#1{#1}\fi

\bibitem[{{Baraffe} {et~al.}(2003){Baraffe}, {Chabrier}, {Barman}, {Allard}, \&
  {Hauschildt}}]{baraffe2003}
{Baraffe}, I., {Chabrier}, G., {Barman}, T.~S., {Allard}, F., \& {Hauschildt},
  P.~H. 2003, \aap, 402, 701

\bibitem[{{Beuzit} {et~al.}(2004){Beuzit}, {S{\'e}gransan}, {Forveille},
  {Udry}, {Delfosse}, {Mayor}, {Perrier}, {Hainaut}, {Roddier}, {Roddier}, \&
  {Mart{\'{\i}}n}}]{beuzit2004}
{Beuzit}, J.-L., {S{\'e}gransan}, D., {Forveille}, T., {et~al.} 2004, \aap,
  425, 997

\bibitem[{{Biller} {et~al.}(2006){Biller}, {Kasper}, {Close}, {Brandner}, \&
  {Kellner}}]{biller2006}
{Biller}, B.~A., {Kasper}, M., {Close}, L.~M., {Brandner}, W., \& {Kellner}, S.
  2006, \apjl, 641, L141

\bibitem[{{Chabrier} {et~al.}(2000){Chabrier}, {Baraffe}, {Allard}, \&
  {Hauschildt}}]{chabrier2000}
{Chabrier}, G., {Baraffe}, I., {Allard}, F., \& {Hauschildt}, P. 2000, \apj,
  542, 464

\bibitem[{{Deacon} {et~al.}(2005){Deacon}, {Hambly}, {Henry}, {Subasavage},
  {Brown}, \& {Jao}}]{deacon2005}
{Deacon}, N., {Hambly}, N., {Henry}, T., {et~al.} 2005, \aj, 129, 409

\bibitem[{{Delfosse} {et~al.}(1999){Delfosse}, {Forveille}, {Beuzit}, {Udry},
  {Mayor}, \& {Perrier}}]{delfosse1999}
{Delfosse}, X., {Forveille}, T., {Beuzit}, J.-L., {et~al.} 1999, \aap, 344, 897

\bibitem[{{Delfosse} {et~al.}(2000){Delfosse}, {Forveille}, {S{\'e}gransan},
  {Beuzit}, {Udry}, {Perrier}, \& {Mayor}}]{delfosse2000}
{Delfosse}, X., {Forveille}, T., {S{\'e}gransan}, D., {et~al.} 2000, \aap, 364,
  217

\bibitem[{{Devillard}(1997)}]{devillard1997}
{Devillard}, N. 1997, The Messenger, 87, 19

\bibitem[{{Duquennoy} \& {Mayor}(1991)}]{duquennoy1991}
{Duquennoy}, A. \& {Mayor}, M. 1991, \aap, 248, 485

\bibitem[{{Halbwachs} {et~al.}(2003){Halbwachs}, {Mayor}, {Udry}, \&
  {Arenou}}]{halbwachs2003}
{Halbwachs}, J.~L., {Mayor}, M., {Udry}, S., \& {Arenou}, F. 2003, \aap, 397,
  159

\bibitem[{{Hambly} {et~al.}(2004){Hambly}, {Henry}, {Subasavage}, {Brown}, \&
  {Jao}}]{hambly2004}
{Hambly}, N., {Henry}, T., {Subasavage}, J., {Brown}, M., \& {Jao}, W.-C. 2004,
  \aj, 128, 437

\bibitem[{{Hawley} {et~al.}(1996){Hawley}, {Gizis}, \& {Reid}}]{hawley1996}
{Hawley}, S., {Gizis}, J., \& {Reid}, I. 1996, \aj, 112, 2799

\bibitem[{{Henry} {et~al.}(2004){Henry}, {Subasavage}, {Brown}, {Beaulieu},
  {Jao}, \& {Hambly}}]{henry2004}
{Henry}, T., {Subasavage}, J., {Brown}, M., {et~al.} 2004, \aj, 128, 2460

\bibitem[{{Jeffries} \& {Maxted}(2005)}]{jeffries2005}
{Jeffries}, R. \& {Maxted}, P. 2005, Astron. Notes, 326, 944

\bibitem[{{Knapp} {et~al.}(2004){Knapp}, {Leggett}, {Fan}, {Marley}, {Geballe},
  {Golimowski}, {Finkbeiner}, {Gunn}, {Hennawi}, {Ivezi{\'c}}, {Lupton},
  {Schlegel}, {Strauss}, {Tsvetanov}, {Chiu}, {Hoversten}, {Glazebrook},
  {Zheng}, {Hendrickson}, {Williams}, {Uomoto}, {Vrba}, {Henden}, {Luginbuhl},
  {Guetter}, {Munn}, {Canzian}, {Schneider}, \& {Brinkmann}}]{knapp2004}
{Knapp}, G.~R., {Leggett}, S.~K., {Fan}, X., {et~al.} 2004, \aj, 127, 3553

\bibitem[{{Lagrange} {et~al.}(2003){Lagrange}, {Chauvin}, {Fusco}, {Gendron},
  {Rouan}, {Hartung}, {Lacombe}, {Mouillet}, {Rousset}, {Drossart}, {Lenzen},
  {Moutou}, {Brandner}, {Hubin}, {Clenet}, {Stolte}, {Schoedel}, {Zins}, \&
  {Spyromilio}}]{lagrange2003}
{Lagrange}, A.-M., {Chauvin}, G., {Fusco}, T., {et~al.} 2003, in Proc. SPIE,
  4841, 860

\bibitem[{{Leggett} {et~al.}(2000){Leggett}, {Allard}, {Dahn}, {Hauschildt},
  {Kerr}, \& {Rayner}}]{leggett2000}
{Leggett}, S., {Allard}, F., {Dahn}, C., {et~al.} 2000, \apj, 535, 965

\bibitem[{{Leggett} {et~al.}(2002){Leggett}, {Golimowski}, {Fan}, {Geballe},
  {Knapp}, {Brinkmann}, {Csabai}, {Gunn}, {Hawley}, {Henry}, {Hindsley},
  {Ivezi{\'c}}, {Lupton}, {Pier}, {Schneider}, {Smith}, {Strauss}, {Uomoto}, \&
  {York}}]{leggett2002}
{Leggett}, S., {Golimowski}, D., {Fan}, X., {et~al.} 2002, \apj, 564, 452

\bibitem[{{Lenzen} {et~al.}(2003){Lenzen}, {Hartung}, {Brandner}, {Finger},
  {Hubin}, {Lacombe}, {Lagrange}, {Lehnert}, {Moorwood}, \&
  {Mouillet}}]{lenzen2003}
{Lenzen}, R., {Hartung}, M., {Brandner}, W., {et~al.} 2003, in Proc. SPIE,
  4841, 944

\bibitem[{{Luyten}(1979)}]{luyten1979}
{Luyten}, W.~J. 1979, LHS catalogue

\bibitem[{{McCaughrean} \& {Stauffer}(1994)}]{mccaughrean1994}
{McCaughrean}, M.~J. \& {Stauffer}, J.~R. 1994, \aj, 108, 1382

\bibitem[{{Patterson} {et~al.}(1998){Patterson}, {Ianna}, \&
  {Begam}}]{patterson1998}
{Patterson}, R., {Ianna}, P., \& {Begam}, M. 1998, \aj, 115, 1648

\bibitem[{{Reid} {et~al.}(2003){Reid}, {Cruz}, {Allen}, {Mungall}, {Kilkenny},
  {Liebert}, {Hawley}, {Fraser}, {Covey}, \& {Lowrance}}]{reid2003}
{Reid}, I., {Cruz}, K., {Allen}, P., {et~al.} 2003, \aj, 126, 3007

\bibitem[{{Reid} {et~al.}(2002){Reid}, {Kilkenny}, \& {Cruz}}]{reid2002}
{Reid}, I., {Kilkenny}, D., \& {Cruz}, K. 2002, \aj, 123, 2822

\bibitem[{{Rigaut} {et~al.}(1998){Rigaut}, {Salmon}, {Arsenault}, {Thomas},
  {Lai}, {Rouan}, {V{\'e}ran}, {Gigan}, {Crampton}, {Fletcher}, {Stilburn},
  {Boyer}, \& {Jagourel}}]{rigaut1998}
{Rigaut}, F., {Salmon}, D., {Arsenault}, R., {et~al.} 1998, \pasp, 110, 152

\bibitem[{{Rousset} {et~al.}(2003){Rousset}, {Lacombe}, {Puget}, {Hubin},
  {Gendron}, {Fusco}, {Arsenault}, {Charton}, {Feautrier}, {Gigan}, {Kern},
  {Lagrange}, {Madec}, {Mouillet}, {Rabaud}, {Rabou}, {Stadler}, \&
  {Zins}}]{rousset2003}
{Rousset}, G., {Lacombe}, F., {Puget}, P., {et~al.} 2003, in Proc. SPIE, 4839,
  140

\bibitem[{{Salim} \& {Gould}(2003)}]{salim2003}
{Salim}, S. \& {Gould}, A. 2003, \apj, 582, 1011

\bibitem[{{Scholz} {et~al.}(2005){Scholz}, {Meusinger}, \&
  {Jahrei{\ss}}}]{scholz2005}
{Scholz}, R.-D., {Meusinger}, H., \& {Jahrei{\ss}}, H. 2005, \aap, 442, 211

\bibitem[{{Siegler} {et~al.}(2005){Siegler}, {Close}, {Cruz}, {Mart{\'{\i}}n},
  \& {Reid}}]{siegler2005}
{Siegler}, N., {Close}, L., {Cruz}, K., {Mart{\'{\i}}n}, E., \& {Reid}, I.
  2005, \apj, 621, 1023

\bibitem[{{Skrutskie} {et~al.}(2006){Skrutskie}, {Cutri}, {Stiening},
  {Weinberg}, {Schneider}, {Carpenter}, {Beichman}, {Capps}, {Chester},
  {Elias}, {Huchra}, {Liebert}, {Lonsdale}, {Monet}, {Price}, {Seitzer},
  {Jarrett}, {Kirkpatrick}, {Gizis}, {Howard}, {Evans}, {Fowler}, {Fullmer},
  {Hurt}, {Light}, {Kopan}, {Marsh}, {McCallon}, {Tam}, {Van Dyk}, \&
  {Wheelock}}]{skrutskie2006}
{Skrutskie}, M., {Cutri}, R., {Stiening}, R., {et~al.} 2006, \aj, 131, 1163

\bibitem[{{V{\'e}ran} {et~al.}(1999){V{\'e}ran}, {Beuzit}, \&
  {Chaytor}}]{veran1999}
{V{\'e}ran}, J.-P., {Beuzit}, J.-L., \& {Chaytor}, D. 1999, in Astronomy with
  adaptive optics : present results and future programs, ed. D.~{Bonaccini},
  691

\bibitem[{{V{\'e}ran} {et~al.}(1997){V{\'e}ran}, {Rigaut}, {Ma{\^\i}tre}, \&
  {Rouan}}]{veran1997}
{V{\'e}ran}, J.-P., {Rigaut}, F., {Ma{\^\i}tre}, H., \& {Rouan}, D. 1997,
  Optical Society of America Journal A, 14, 3057

\bibitem[{{Wroblewski} \& {Torres}(1991)}]{wroblewski1991}
{Wroblewski}, H. \& {Torres}, C. 1991, \aaps, 91, 129

\end{thebibliography}

\end{document}